# SLO-Aware Compute Resource Allocation for Prefill-Decode Disaggregated LLM Inference

Luchang Li*, Dongfang Li, Bozhao Gong, Yu Zhang
Kingsoft Cloud

**Abstract**

Prefill-Decode (P/D) disaggregation has emerged as a widely adopted optimization strategy for Large Language Model (LLM) inference. However, there currently exists no well-established methodology for determining the optimal number of P/D hardware resources, subject to constraints on total throughput, service level objectives (SLOs), and request characteristics—specifically input and output lengths. To address this gap, we propose a hybrid approach that combines theoretical modeling with empirical benchmarking. First, we present a theoretical model for calculating P/D resource counts, which is based on total throughput requirements, request input and output lengths, as well as prefill and decode throughput. Then, to obtain the actual prefill and decode throughput under SLO constraints, we model the prefill process using M/M/1 queuing theory, deriving the achieved prefill throughput from the benchmarked maximum prefill throughput and Time-To-First-Token (TTFT). For the decode phase, we determine the decode batch sizes that meet Time-Per-Output-Token (TPOT) requirements and obtain the corresponding decode throughput through empirical measurements. Our experimental results demonstrate that the proposed method can accurately predict optimal P/D resource allocation in real-world LLM inference scenarios.

## Introduction

Large Language Model (LLM) have demonstrated remarkable capabilities across a wide range of tasks[9, 16, 20, 24], driving a surge in demand for inference serving. Deploying LLM with high performance while satisfying Service Level Objectives (SLOs) is critical to ensuring user experience and reducing inference costs. Traditionally, LLM inference is deployed in non-disaggregated manner, where the same set of GPU resources handles compute-bound prefill and memory-bandwidth-bound decode sequentially. The two distinct phases interfere with each other, making it difficult to simultaneously optimize the Time To First Token (TTFT) and Time Per Output Token (TPOT) metrics[12, 23]. It also hinders the independent optimization of the two phases, for instance, employing different hardware[2, 21], diverse parallelism strategies, and other approaches to improve throughput.

To mitigate these interference issues, Prefill-Decode (P/D) disaggregation has emerged as a widely adopted optimization paradigm[2, 5, 12, 21, 23]. By decoupling the two phases into separate prefill and decode instances, and transferring the KV cache between them, both SLOs and throughput of the two phases can be independently optimized. Existing frameworks like vLLM[7], SGLang[22], TensorRT-LLM[25] etc. have successfully implemented the mechanisms for P/D disaggregation. However, a critical operational challenge remains: how to determine the optimal provision of hardware resources (i.e., the specific count of prefill vs. decode GPUs) for a given workload. Improper P/D resource allocation either results in resource underutilization or fails to meet the SLO requirements.
Currently, the industry lacks a well established methodology to calculate P/D resource counts precisely. SGLang deployed DeepSeek V3.1 on $P_3D_9$ instances[26], but the specific P/D instance ratio has not been disclosed. NVIDIA's AIConfigurator[19] employs a search-based approach to derive optimal deployment strategies for P/D under specific user SLO requirements, such as parameter settings for Tensor Parallelism (TP), attention Data Parallelism (DP), Expert Parallelism (EP), among others. Nevertheless, it still fails to provide a comprehensive resource allocation method for P/D under user's throughput and SLO requirements. Some other works[8, 13, 17] also schedule PD resources, but they do not fully take into account the requirements of SLOs or the actual prefill and decode throughput of the system.
To address this gap, we propose a novel approach that integrates theoretical modeling with empirical benchmarking to precisely determine the optimal P/D resource allocation. Our approach does not targeting the optimal deployment of individual P/D instances. Instead, under a pre-determined deployment, it determines the resource quantity allocated to P/D according to user-specified throughput and SLO requirements.
The main contributions of our paper include:

- We establish a theoretical model that calculates the count of P/D instances based on total throughput, SLOs requirements, request input/output length, and the achievable throughput of P/D stages.
- To determine the achievable prefill throughput under TTFT constraints, we model the prefill process by M/M/1 queuing theory. This allows us to derive the effective prefill throughput that satisfies TTFT requirements, constrained by the benchmarked maximum prefill throughput.
- Simultaneously, we employ empirical measurements to identify the maximum decode batch size that satisfies TPOT requirements, thereby deriving the effective decode throughput .
- We demonstrate that our method can accurately predict the optimal P/D resource allocation in real-world inference scenarios, ensuring both cost-efficiency and strict SLO compliance.

* Corresponding author: liluchang@kingsoft.com

# Method

## PD hardware resource quantity calculation

First, the user-specified total throughput refers to the total number of input and output tokens processed per second. Therefore, the total throughput $TP_{total}$ is calculated as follows:

$$TP_{total} = \frac{N_{req} * (L_{in} + L_{out})}{T_{total}} \quad (1)$$

Here, $N_{req}$ denotes the number of requests, while $L_{in}$ and $L_{out}$ represent the average input and output sequence lengths per request, respectively. $T_{total}$ refers to the total time required to complete the prefill and decode computations for all requests.
In a P/D disaggregated system, the input processing and output generation of requests are handled by the prefill and decode components, respectively. For these two phases, prefill throughput $TP_{prefill}$ and decode throughput $TP_{decode}$ are defined as the number of input tokens processed and output tokens generated per second by a single prefill or decode instance, respectively. Thus, we obtain the processing times for the prefill and decode phases as follows:

$$T_{prefill} = \frac{N_{req} * L_{in}}{TP_{prefill} * N_{prefill}} \quad (2)$$

$$T_{decode} = \frac{N_{req} * L_{out}}{TP_{decode} * N_{decode}} \quad (3)$$

Here, $N_{prefill}$ and $N_{decode}$ denote the number of prefill and decode instances, respectively. The P/D disaggregated system adopts pipelined inference, so the total computation time is the maximum of the prefill and decode computation times. To avoid idling of either the prefill or decode components, the computation times of the two phases should be equal, i.e.:

$$T_{total} = T_{prefill} = T_{decode} \quad (4)$$

Substituting Formula (2) and Formula (3) into Formula (1), we can derive the number of prefill and decode instances, respectively:

$$N_{prefill} = \frac{TP_{total} * L_{in}}{(L_{in} + L_{out}) * TP_{prefill}} \quad (5)$$

$$N_{decode} = \frac{TP_{total} * L_{out}}{(L_{in} + L_{out}) * TP_{decode}} \quad (6)$$

Dividing Formula (5) by Formula (6) yields the P/D ratio without requiring the total throughput:

$$R_{P/D} = \frac{L_{in} * TP_{decode}}{L_{out} * TP_{prefill}} \quad (7)$$

In these formulas, the total throughput, input length, and output length are known user-specified requirements. To determine the number and ratio of P/D instances, it is necessary to first obtain the throughput of individual prefill and decode instances. The following sections describe how to calculate the corresponding prefill and decode throughput under specific SLO requirements and given request input and output lengths.

## Acquire prefill throughput under TTFT constraints

In this chapter, we introduce how to obtain the relationship between the actual prefill throughput that meets the target TTFT and the empirically measured maximum prefill throughput. First, for the prefill phase, given specific hardware and deployment parameters (e.g., parallelism like TP and EP, chunked prefill size[1]), and request input length, we can measure the maximum prefill throughput $\widehat{TP}_{prefill}$ by fully utilizing prefill GPUs without idle time. However, different TTFT requirements lead to different prefill computation utilization rates, resulting in distinct effective prefill throughputs.

In a P/D disaggregated deployment, TTFT includes not only the request queuing time $T_{prefill_queuing}$ and computation time $T_{prefill_computation}$ in the prefill phase, but also certain overhead time $T_{overhead}$. This $T_{overhead}$ consists of the input and output transmission time of requests between the user terminal and the GPU server, as well as the transmission time of the KV cache between the P and D instances. The $T_{overhead}$ can typically be pre-determined. Therefore, TTFT is calculated as:

$$\mathrm{TTFT} = \mathrm{T_{prefill_queuing}} + \mathrm{T_{prefill_computation}} + \mathrm{T_{overhead}} \quad (8)$$

Second, we model the request queuing and computation process of a single prefill instance as an M/M/1 queuing problem[3, 6]. Note that when DP[27] is enabled in prefill deployment, each DP group processes requests independently; thus, the M/M/1 model should be applied to each DP group separately. For the sake of simplicity, this paper only discusses the scenario where DP is disabled.

Based on the M/M/1 queuing model, the service rate μ is defined as the ratio of the maximum prefill throughput to the input sequence length:

$$\mu = \frac{\widehat{\mathrm{TP_{prefill}}}}{\mathrm{L_{in}}} \quad (9)$$

The arrival rate λ is defined as the actual request rate, which must be smaller than μ. According to queuing theory, the system utilization ρ is calculated as:

$$\rho = \frac{\lambda}{\mu} \quad (10)$$

We then define the actually achieved prefill throughput based on the utilization:

$$\mathrm{TP_{prefill}} = \widehat{\mathrm{TP_{prefill}}} * \rho \quad (11)$$

According to the M/M/1 queuing model, the formula for the sum of the prefill queuing and computation time is:

$$\mathrm{T_s} = \mathrm{T_{prefill_queuing}} + \mathrm{T_{prefill_computation}} = \mathrm{TTFT} - \mathrm{T_{overhead}} = \frac{1}{\mu - \lambda} \quad (12)$$

Finally, based on the above derivations, we can derive the actual throughput from the target TTFT and the maximum prefill throughput as follows:

$$\mathrm{TP_{prefill}} = \widehat{\mathrm{TP_{prefill}}} - \frac{\mathrm{L_{in}}}{\mathrm{TTFT} - \mathrm{T_{overhead}}} \quad (13)$$

Two key insights can be drawn from this formula: first, the lower the target TTFT value, the lower the actual prefill throughput that can be achieved; second, for the same target TTFT, the higher the maximum prefill throughput, the higher the system utilization rate that can be achieved.

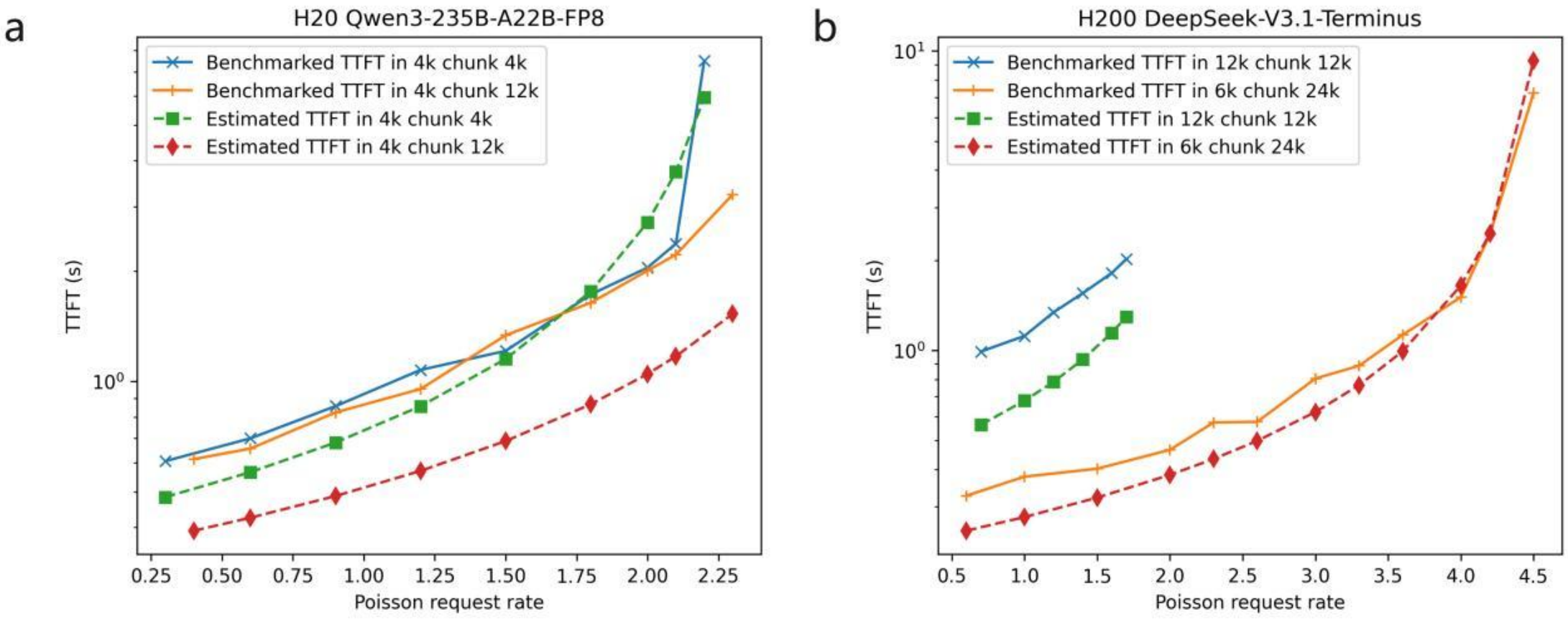

Figure 1 Comparison between benchmarked TTFT and M/M/1 queuing theory-predicted TTFT. Note that the predicted TTFT in this figure does not include the KV cache transmission time from prefill to decode. In this figure, in means input sequence length, chunk means chunked

prefill size.

We verify the correctness of this theory by comparing the actually measured TTFT with the value predicted by Formula (12). We deploy one P instance and one D instance using 1 NVIDIA H20 node with the Qwen3-235B-A22B-FP8 model, where each instance occupies 4 GPUs. In addition, we deploy one P instance and one D instance using 2 NVIDIA H200 nodes with the DeepSeek-V3.1-Terminus model, where each instance occupies 1 node with 8 GPUs. We note that the prefill deployment has a parameter called chunked prefill size. Within a certain range, the larger the chunked prefill size, the higher the peak prefill throughput, until saturation is reached. The M/M/1 queuing model requires requests to be processed sequentially, which holds when the request length is greater than or equal to the chunked prefill size. When the chunked prefill size is much larger than the request length, multiple requests in the queue may be batched and processed together, which does not fully align with the M/M/1 queuing theory.

To fully comply with the assumptions of the M/M/1 queuing theory, we first evaluated the case where the chunked prefill size equals the request input length. We measured the variation in request TTFT with request rate for an input length of 4096 on the H20 deployment and for an input length of 12288 on the H200 deployment. We then evaluated the corresponding maximum performance and predicted the theoretical TTFT values using Formula (12); the results are shown in Figure 1. As observed from the figure, the predicted TTFT exhibits perfect trend consistency with the measured results, apart from a small gap. This gap is primarily attributed to the KV cache transmission time. Note that the predicted TTFT in Figure 1 does not include the KV cache transmission time from prefill to decode, and the KV cache transmission time increases with longer request input lengths.

As shown in Figure 1, we also evaluated scenarios where the chunked prefill size is much larger than the request length. For example, we evaluated an input length of 4096 with a chunked prefill size of 12288, and an input length of 6144 with a chunked prefill size of 24576. We found that the results based on our method still serve as a good approximation.

In summary, our method first requires evaluating the maximum prefill throughput of the deployed prefill instance under non-idle conditions. We then model the prefill process as an M/M/1 queuing problem and derive Formula (13), which calculates the actually achieved prefill throughput using the maximum prefill throughput, required TTFT, and request input length. Note that our method does not account for prefix caching[4, 10, 11, 18]. If the prefill computation time is proportional to the length of the non-prefix-cache-hit portion, it is only necessary to replace the input length in this paper with the input length that does not hit the KV cache.

## Acquire decode throughput under TPOT constraints

In this section, we describe how to obtain the actual decode throughput that meets the target TPOT. During the decoding phase, the LLM engine leverages continuous batching to compute multiple requests in parallel. We find that both decode TPOT and decode throughput are positively correlated with the decoding batch size. Specifically, a larger batch size leads to higher decode throughput but also increases TPOT. Therefore, to achieve the decode throughput that satisfies the target TPOT, we only need to first benchmark the curves of actual decode throughput and TPOT against the decoding batch size, and then determine the decode throughput that meets the TPOT requirement.

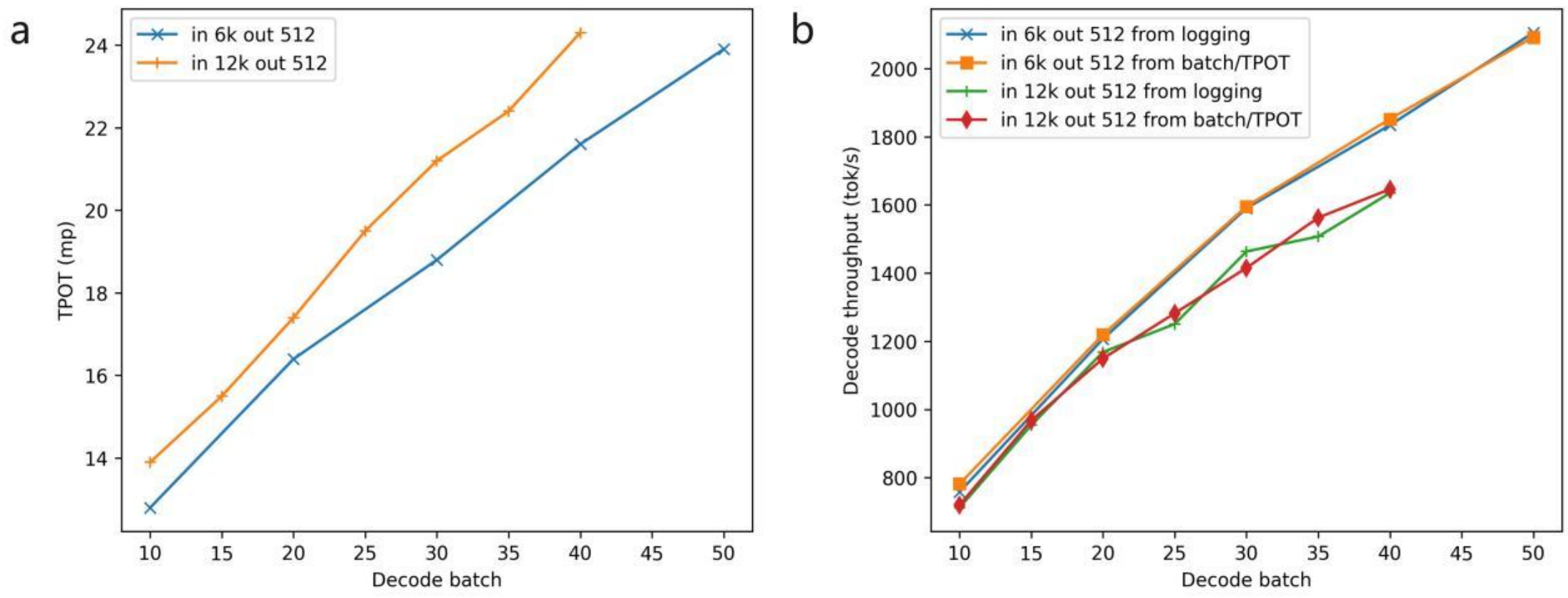

Figure 2 Benchmarked curves of TOPT and decode throughput varying with batch size for the DeepSeek-V3.1-Terminus model deployed on an H200 node. Note the multi-token prediction was enabled in the benchmark.

As shown in Figure 2, we evaluated the variation curves of TPOT and decoding throughput across different decoding batch sizes for the DeepSeek-V3.1-Terminus model deployed with 8 NVIDIA H200 GPUs. We tested two input sequence lengths (6144 and 12288) and a fixed output length of 512. TPOT is obtained from the results of the benchmark tool EvalScope[15], while decoding throughput is measured via two independent approaches: one extracted from runtime logs printed by the SGLang inference engine, and the other computed as the decoding batch size divided by the corresponding TPOT. We observe that the throughput results from these two methods are highly consistent. Thus, we can easily obtain decode throughput by benchmarking TPOT and the corresponding batch size without recording and parsing runtime logs. Notably, this throughput calculation method requires that the decoding computation batch size is approximately consistent with the benchmarked concurrency size.
In summary, to attain the decoding throughput that meets the TPOT requirement, we first characterize the curve of decoding TPOT as a function of batch size under realistic user input – output workloads. We then determine the batch size that satisfies the target TPOT constraint, and derive the corresponding decoding throughput by dividing this valid batch size by the associated TPOT.

# Evaluation

We validated our method through a realistic inference scenario. The user requirements are specified as follows:

- LLM model: DeepSeek-V3.1-Terminus
- TTFT: 2 s
- TPOT: 20 ms
- Mean input sequence length: 6144
- Mean output sequence length: 512
- Total Throughput: 5 million tokens per minute (M TPM)

The hardware and software employed are listed as follows:

- Deployment hardware: NVIDIA H200 GPUs

- Inference engine: SGLang v0.5.8
- Benchmark tool: EvalScope v1.4.2

To meet the SLO constraints, we deployed the prefill instances using a combination of TP and EP, with a chunked prefill size of 24576; the two-batch overlap optimization was disabled. For the decode instances, we only used TP and disabled both DP and EP. Each P/D instance ran on an independent node, and Multi-Token Prediction (MTP) was enabled for all instances.

We evaluated that the maximum prefill throughput for this input sequence length is 28300 tokens/s. Considering a 100 ms KV cache transfer time (i.e., $T_{overhead}$ = 100 ms), we derived an effective prefill throughput of approximately 25000 tokens/s using Formula (13).

From the benchmarked curves of TPOT and decode throughput versus decode batch size, as illustrated in Figure 2, we obtained a decode throughput of approximately 1700 tokens/s that satisfies the TPOT constraint. Based on Formula (7), we calculated a P:D ratio of 0.82:1. Using Formulas (5) and (6), we determined that three prefill instances and four decode instances are required to meet the total throughput demand. In this paper, we use the notation mPnD to denote a deployment with m prefill instances and n decode instances. Therefore, we propose a 3P4D deployment for this scenario.

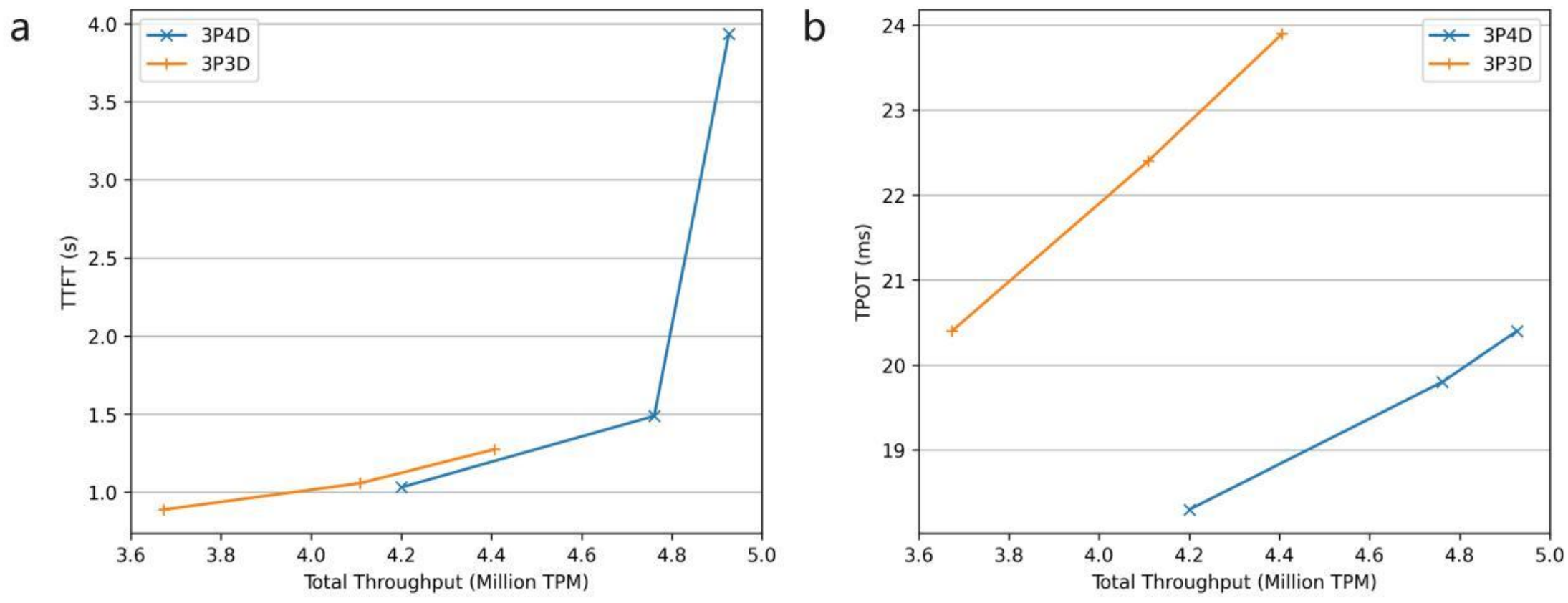

Figure 3 Curves of TTFT and TPOT varying with total throughput under different P/D resource counts.

We evaluated the curves of TTFT and TPOT as functions of total throughput for the 3P4D deployment. We also evaluated the 3P3D scenario for comparison. The results are shown in Figure 3, from which we can observe that when deployed with 3P4D, both the 2 s TTFT and 20 ms SLO thresholds are achieved simultaneously at approximately 4.8 M TPM. This demonstrates that the P/D ratio calculated by our method is well-balanced, and the achieved total throughput is very close to the 5 M TPM target value. In contrast, when using the 3P3D deployment, the SLO threshold can only be met at approximately 3.6 M TPM, which is mainly constrained by TPOT, while TTFT still has a considerable margin. Under this SLO, the average throughput per node for 3P4D is 0.69 M TPM, compared with 0.6 M TPM for 3P3D, indicating that the deployment efficiency calculated by our method is higher.

# Conclusion

In this paper, we have addressed the critical challenge of computing resource allocation for P/D disaggregated LLM inference. We propose a hybrid methodology that integrates theoretical modeling with empirical benchmarking. First, we derive a formula to determine the required number of P/D resources based on total throughput requirements, SLO constraints, request input/output lengths, and P/D throughput. In this formula, most parameters are user-specified known values, whereas the actually achievable throughput of P/D deployments remains unclear. Second, we employ M/M/1 queuing theory to model the prefill phase, which enables us to derive the effective throughput that meets the TTFT constraint. For the decode phase, we conduct empirical measurements to identify the optimal batch size that satisfies the TPOT requirement, thereby ensuring maximum resource utilization while avoiding violations of performance guarantees. Finally, we validate our method through real-world inference scenarios, demonstrating its capability to accurately predict P/D resource counts and thus maximize the efficiency of P/D resource utilization under given SLO constraints.

In the future, our method could be integrated with approaches such as AIConfigurator, which estimates the optimal deployment of individual P/D instances and employs simulation-based methods to predict P/D throughput. Furthermore, our method has the potential to be generalized to multimodal EPD separation systems[14], enabling the determination of resource counts for the three independently deployed components.